%% file: 0_main.tex
\documentclass[runningheads]{llncs}
\usepackage[T1]{fontenc}
\usepackage{subfig} 
\usepackage{graphicx}
\usepackage{todonotes}
\usepackage{hyperref}
\usepackage{booktabs}
\usepackage{appendix}
\usepackage{multirow}
\usepackage{amsmath}
\usepackage{amssymb}
\usepackage{amsfonts}

\begin{document}
\title{Diffusion Models for Unsupervised Anomaly Detection in Fetal
Brain Ultrasound}
\titlerunning{Diffusion Models for UAD in Fetal Brain US}
%
\author{Hanna Mykula\inst{1} \and
Lisa Gasser\inst{3} \and 
Silvia Lobmaier\inst{3} \and 
Julia A. Schnabel\inst{1 2 4} \and
Veronika Zimmer$^*$\inst{1} \and 
Cosmin I. Bercea$^*$\inst{1 2}}
\authorrunning{Hanna M. et al.}
%
\institute{Technical University of Munich, Germany \and
Helmholtz AI and Helmholtz Center Munich, Germany \and 
Klinikum rechts der Isar, Technical University of Munich, Germany \and 
King’s College London, UK
}
\def\thefootnote{*}\footnotetext{Contributed equally as senior authors.}

%
\maketitle              
\begin{abstract}
Ultrasonography is an essential tool in mid-pregnancy for assessing fetal development, appreciated for its non-invasive and real-time imaging capabilities. Yet, the interpretation of ultrasound images is often complicated by acoustic shadows, speckle noise, and other artifacts that obscure crucial diagnostic details. To address these challenges, our study presents a novel unsupervised anomaly detection framework specifically designed for fetal ultrasound imaging. This framework incorporates gestational age filtering, precise identification of fetal standard planes, and targeted segmentation of brain regions to enhance diagnostic accuracy. Furthermore, we introduce the use of denoising diffusion probabilistic models in this context, marking a significant innovation in detecting previously unrecognized anomalies. We rigorously evaluated the framework using various diffusion-based anomaly detection methods, noise types, and noise levels. Notably, AutoDDPM emerged as the most effective, achieving an area under the precision-recall curve of 79.8\% in detecting anomalies. This advancement holds promise for improving the tools available for nuanced and effective prenatal diagnostics.
\keywords{Fetal Ultrasound Screening \and Medical Imaging}

\begin{figure}[b!]
  \centering
  \includegraphics[width=0.8\textwidth]{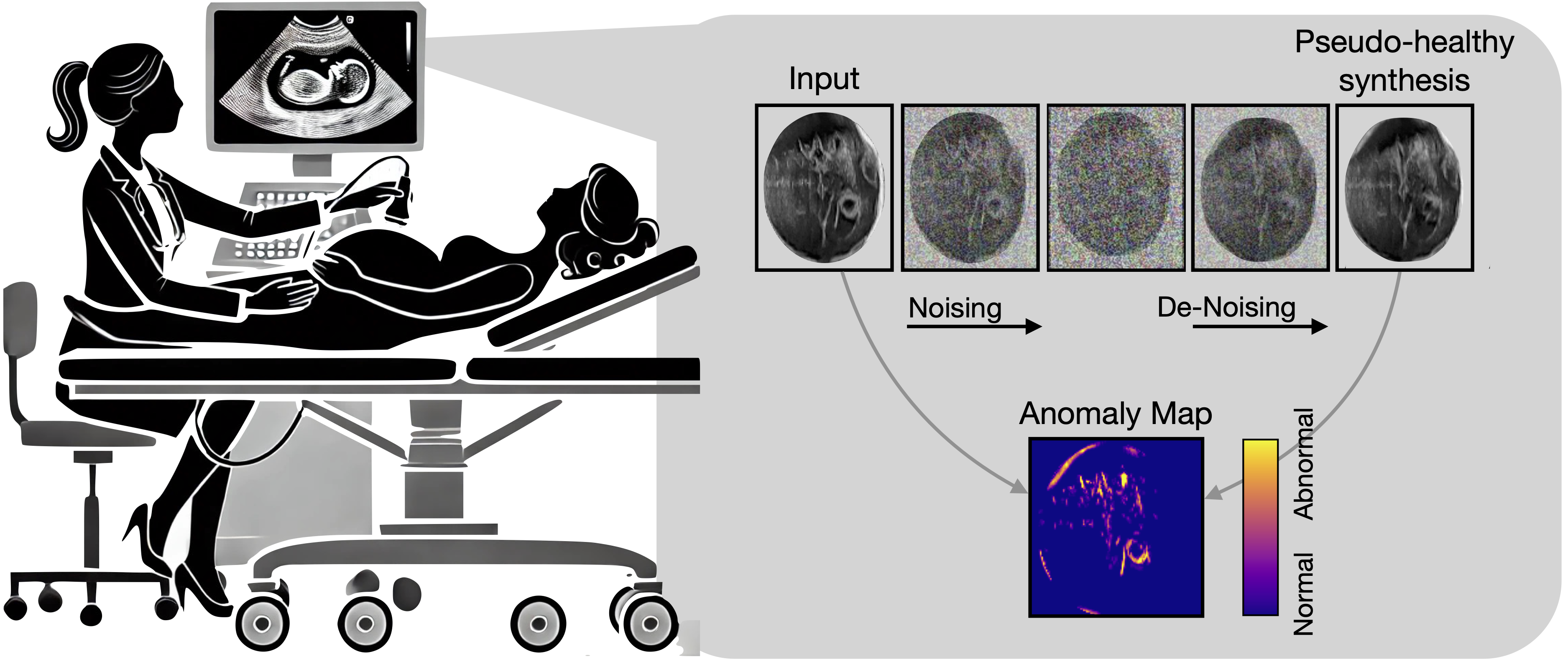}
  \caption{Workflow for screening and monitoring fetal brain anomalies in ultrasound images during pregnancy. Diffusion models generate pseudo-healthy synthesis to effectively identify potential developmental anomalies.}
\end{figure}

\end{abstract}

\input{sections/1_intro}
\input{sections/2_related_work}
\input{sections/3_background}
\input{sections/4_methods_datasets}

\input{sections/5_results_experiments}
\input{sections/6_conclusion}

%
%
%
%
\bibliographystyle{splncs04}
\bibliography{0_main}

\end{document}


\title{Diffusion Models for Unsupervised Anomaly Detection in Fetal
Brain Ultrasound}
\titlerunning{Diffusion Models for UAD in Fetal Brain US}
%
\author{Hanna Mykula\inst{1} \and
Veronika Zimmer\inst{1} \and 
Lisa Gasser\inst{3} \and 
Silvia Lobmaier\inst{3} \and 
Julia A. Schnabel$^*$\inst{1 2 4} \and
Cosmin I. Bercea$^*$\inst{1 2}}
\authorrunning{Hanna M. et al.}
%
\institute{Technical University Munich, Germany \and
Helmholtz AI and Helmholtz Center Munich, Germany \and 
Klinikum rechts der Isar, Technical University of Munich, Germany \and 
King’s College London, UK
}
\def\thefootnote{*}\footnotetext{Contributed equally as senior authors.}

%
\maketitle              

\appendix
\input{sections/99_apendix}

%% file: sections/1_intro.tex
\section{Introduction}

Ultrasonography (US) is an indispensable tool in prenatal care, widely used for monitoring fetal development due to its safety, real-time imaging capabilities, and cost-effectiveness~\cite{PrenatalImaging}. Particularly, the mid-pregnancy US scan at around 22 weeks is crucial for assessing fetal growth and identifying potential anomalies, including those affecting the brain. However, interpreting US images remains challenging due to artifacts such as acoustic shadows, speckle noise, and motion blurring. These issues arise from the complex interactions between US waves and biological tissues, which can obscure critical diagnostic details and complicate both manual and automated analyses~\cite{Meng_2019}.

The application of deep learning (DL) in fetal US image analysis has shown significant promise, enhancing the ability of clinicians to detect anomalies~\cite{Fiorentino_2023}. However, supervised DL methods face considerable challenges due to the anatomical diversity and the imbalance between healthy and anomalous samples~\cite{Pang_2021,Zhang_2022}. Given these limitations, unsupervised anomaly detection (UAD) emerges as a viable alternative. UAD methods train exclusively on healthy samples, establishing a baseline for normality without the need for labeled anomaly data. Even though this approach offers better generalization across various pathologies~\cite{Tschuchnig_2022}, it remains largely unexplored for fetal brain US.

In this study, we propose a novel UAD framework aimed at detecting brain anomalies in fetal US images. Our framework incorporates several innovative techniques to enhance the quality and consistency of the US scans. Specifically, we filter by gestational age, detect and localize standard planes using SonoNet~\cite{baumgartner2017sononet}, and remove background noise to focus on the brain region. Furthermore, we pioneer the application of denoising diffusion probabilistic models (DDPM) in this context. DDPMs have shown exceptional performance in capturing complex distributions and generating high-fidelity images, making them suitable for identifying previously unseen anomalies in fetal US scans~\cite{DBLP:journals/corr/abs-2006-11239}.

We thoroughly evaluated our framework using various diffusion-based UAD methods, noise types, and noise levels to understand their performance and robustness. Among these, AutoDDPM~\cite{bercea2023mask} demonstrated superior performance, achieving an area under the precision-recall curve of 79.8\% in anomaly detection. 


%% file: sections/2_related_work.tex
\section{Related work}

This section reviews existing research on anomaly detection (AD) in brain imaging, focusing on two main areas: AD methods for brain MRI and AD methods for fetal US. This overview establishes the context for our novel approach using diffusion-based methods for fetal brain US anomaly detection.\\

\noindent\textbf{Anomaly Detection in Brain MRI.}
Magnetic resonance imaging (MRI) is vital for detecting brain abnormalities, providing detailed tissue contrasts and revealing various pathological changes~\cite{Luo_2023}. Traditional AD methods in brain MRI include Autoencoders~\cite{bercea2023aes,chen2020unsupervised,zimmerer2019unsupervised} and Generative Adversarial Networks~\cite{schlegl2019fanogan,akcay2019ganomaly}, which learn the distribution of healthy anatomy by compressing and decompressing image data. Recently, diffusion models have demonstrated superior performance by offering better mode coverage and sample quality. Denoising diffusion probabilistic models (DDPMs) iteratively learn the data distribution through noising and denoising processes, showing significant success in brain MRI applications~\cite{behrendt2023patched,wolleb2022diffusion,bercea2023mask}. However, their application to fetal brain US remains unexplored.\\

\noindent\textbf{Anomaly Detection in Fetal Images.}
There has been significant focus on other fetal organs, particularly the heart, for detecting anomalies in fetal imaging. In heart imaging, Chotzoglou et al.~\cite{kainz_fetal_heart} proposed an unsupervised approach for detecting Hypoplastic Left Heart Syndrome from fetal US images. Research on AD for fetal brain imaging has also seen notable contributions. FOAC-NET, a supervised convolutional neural network (CNN) architecture, has been developed for detecting fetal organ anomalies in MRI~\cite{10.3389/frai.2022.832485}. In fetal brain US, H. N. Xie et al.~\cite{xie_2020} developed a CNN-based system to classify US images into normal and abnormal categories, achieving high accuracy. However, unsupervised anomaly detection remains largely unexplored in fetal brain US.

%% file: sections/3_background.tex
\section{Methods}
\subsection{Background\label{sec::background}}
Denoising Diffusion Probabilistic Models (DDPMs) use a forward diffusion process, $q(x_t | x_{t-1})$, to incrementally corrupt data from a target distribution, $q(x_0)$, to a normal distribution. A reverse process, $p(x_{t-1} | x_t)$, generates samples by transforming noise back to $q(x_0)$.
The forward process is defined as:
\begin{equation}
q(x_t | x_{t-1}) = \mathcal{N}(x_t | x_{t-1}, \sqrt{1 - \beta_t}, \beta_t I),
\end{equation}
with a variance schedule, $\beta_t \in (0, 1)$, increasing linearly from $\beta_1 = 10^{-4}$ to $\beta_T = 0.02$~\cite{ho2020denoising}.
The reverse generative model, with parameters $\theta$, begins with $x_T \sim \mathcal{N}(0, I)$ and proceeds from $T$ to $1$:
\begin{equation}
p_{\theta}(x_{t-1}|x_t) = \mathcal{N}(x_{t-1} | \mu_{\theta}(x_t, t), \tilde{\beta}_t I),
\end{equation}
where $\tilde{\beta}_t = \frac{1 - \bar{\alpha}_{t-1}}{1 - \bar{\alpha}_t} \beta_t$, $\alpha_t = 1 - \beta_t$ and $\bar{\alpha}_t = \prod_{i=0}^{t} \alpha_i$.
A U-Net~\cite{ronneberger2015unet} is used to learn the noise $\epsilon_{\theta}(x_t, t)$ and approximate $\mu_{\theta}$.
The loss function, $\mathcal{L}_{t}$, targeting the marginal likelihood $p_{\theta}(x_0)$, is:

\begin{equation}
\mathcal{L}_t = D_{KL}(q(x_{t-1}|x_t, x_0) || p_{\theta}(x_{t-1}|x_t)),
\end{equation}
where $D_{KL}$ is the Kullback-Leibler divergence. We use Ho et al.'s simplified objective~\cite{ho2020denoising}:
\begin{equation}
\mathcal{L}_s = \mathbb{E}_{x_0 \sim q(x_0), \epsilon \sim \mathcal{N}(0, I)}\left[ \left\| \epsilon - \epsilon_{\theta}(x_t, t) \right\|^2 \right].
\end{equation}
AnoDDPM \cite{Wyatt_2022_CVPR} leverages DDPMs with either Gaussian or Simplex noise for anomaly detection segmentation. A more recent method proposes a conditional diffusion model to produce more accurate pseudo-healthy counterfactuals, known as AutoDDPM~\cite{bercea2023mask}. 

%% file: sections/4_methods_datasets.tex
\subsection{Fetal Brain UAD Framework}

\begin{figure}[tb!]
  \centering
  \includegraphics[width=\textwidth]{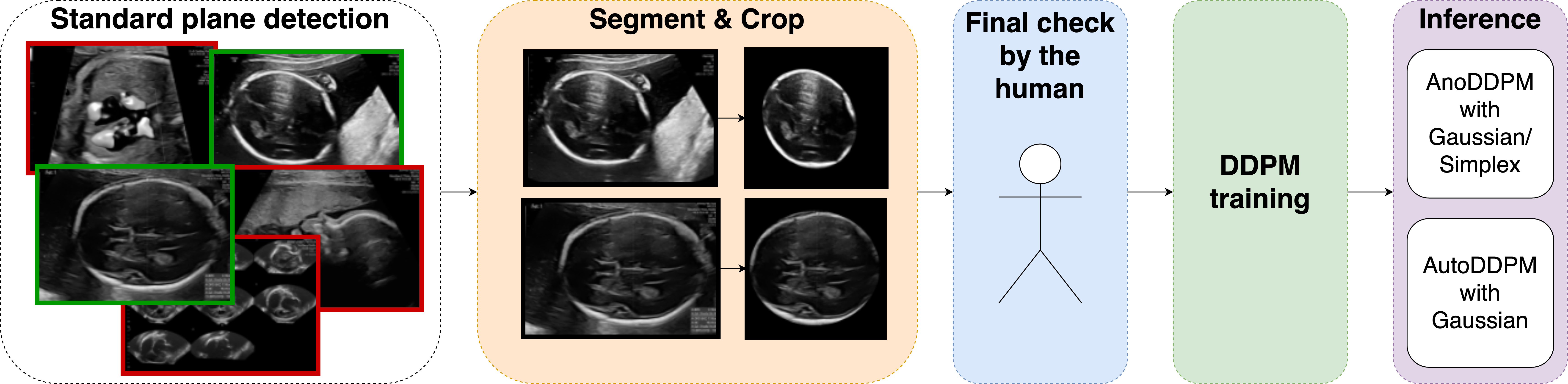}
  \caption{Overview of the proposed framework for detecting brain anomalies in fetal ultrasound scans. The process includes: (1) detecting standard fetal planes (green boxes indicate correct detections, red boxes incorrect) using SonoNet~\cite{baumgartner2017sononet}; (2) segmenting and cropping brain regions from these planes~\cite{biometrics,biometrics_2}; (3) performing a final quality check by a human expert; (4) using the curated dataset to train and evaluate diffusion-based anomaly detection methods; and (5) applying diffusion-based models for anomaly detection through iterative noising and denoising.} \label{fig:framework}
\end{figure}

Selecting US data from the initial dataset requires meticulous attention to ensure the high quality of the training, validation, and testing data. This selection process must meet specific criteria, such as the correct gestational age, the transventricular plane view, and the visibility of key brain structures. To diminish the noise inherent in US images, segmentation into brain and background is necessary. Manually performing this selection process can be extremely time-consuming. Therefore, we propose a semi-automatic data preprocessing pipeline (see Fig. \ref{fig:framework}) that aids in the data curation process, thereby enhancing the quality of the data in the training dataset. Following the construction of the final training, validation, and testing datasets using our semi-automatic pipeline, we train DDPMs to remove artificially added noise and reconstruct pseudo-healthy images. The modular inference setup of our method allows us to test different inference strategies, including AnoDDPM and AutoDDPM, with various noise types such as Simplex and Gaussian.

\subsubsection{Standard Plane Detection.}
We utilize SonoNet~\cite{baumgartner2017sononet} to automatically select images with the correct transventricular plane view of the fetal brain. SonoNet is designed for the real-time detection of fetal standard scan planes in US images and classifies images into one of 13 standard plane categories, including the brain view at the posterior horn of the ventricle (Brain (tv.)). The core of the method is a CNN inspired by the VGG16 architecture. For our experiment, we employed SonoNet-32, equipped with 32 kernels, achieving an F1-Score of 0.798 in plane classification. SonoNet also provides a confidence score for its predictions, enabling a more rigorous selection of candidate images. We eliminate all images that are not labeled as "Brain (tv.)" or have a corresponding confidence score lower than 0.9.

\subsubsection{Brain Segmentation.}
In the second step of our pipeline, we segment and crop all the images to eliminate background noise. We use a probabilistic deep learning approach with a U-Net segmentation network to mask the head from US images. Although in~\cite{biometrics,biometrics_2} ellipses are fitted to the segmented contours for biometric measurements, we only utilize the segmentation model. A manual review is performed to verify the quality of the data that has been automatically selected and preprocessed.

\subsubsection{Diffusion-based Anomaly Detection.}
The final step in our methodology involves the use of DDPMs for anomaly detection. The inference phase of our method is modular, allowing the use of different strategies for anomaly detection. We explore the use of both AnoDDPM and AutoDDPM methods. 

By comparing the performance of AnoDDPM~\cite{Wyatt_2022_CVPR} and AutoDDPM~\cite{bercea2023mask} with different types and levels of noise, we aim to evaluate the effectiveness of different diffusion-based methods for fetal brain US anomaly detection.



%% file: sections/5_results_experiments.tex
\section{Experiments and Results}
\subsection{Experimental Setup}

\noindent{\textbf{Datasets.}} For our experiments, we utilized data from both public~\cite{burgos_artizzu_2020_3904280} and private clinical datasets. The clinical dataset comprised 234 control patients, from which only those within a gestational age range of 19 to 22+6 weeks were selected for inclusion. Consequently, our final dataset for training, validation, and testing included 76 patients from the clinical dataset and 19 from the public dataset, totaling 252 images. For the evaluation of the downstream task, we chose 8 anomalous and 5 healthy control patients from the private clinical dataset, yielding 18 and 12 images, respectively. We adjusted the input pixel values to fall within the range of (0, 1) for Gaussian noise and $(-1, 1)$ for Simplex noise. We normalized the images to the 98th percentile, resized them to a resolution of $128 \times 128$, and applied rotations and horizontal and vertical flips as data augmentation.\\

\noindent{\textbf{Evaluation Metrics.}} To evaluate the performance on healthy scans, we utilized several metrics, including mean absolute error (MAE) as a measure of reconstruction error, structural similarity index (SSIM), and learned perceptual image patch similarity (LPIPS)~\cite{zhang2018unreasonable} for reconstruction accuracy. For anomaly detection performance, we assessed the algorithm’s ability to correctly classify images as either healthy or anomalous using true positives, true negatives, false positives, and false negatives. We calculated the Area Under the Precision-Recall Curve (AUPRC) and the Area Under the Receiver Operating Characteristic (AUROC) curve to provide a comprehensive view of the algorithm’s classification performance.

\subsection{Anomaly Detection Performance}
\begin{table}[tb!]
\caption{Anomaly detection performance. Best results are shown in \textbf{bold} and secondbest are \underline{underlined}.}
\label{tab:classification}
\centering
\begin{tabular}{l|c|c|c|c}
\toprule
\multirow{2}{*}{Method} & \multicolumn{2}{c|}{Healthy} & \multicolumn{2}{c}{Pathological} \\
 & SSIM $\uparrow$  & LPIPS $\downarrow$  & AURPC $\uparrow$ & AUROC $\uparrow$ \\
\midrule
AnoDDPM~\cite{Wyatt_2022_CVPR} with Simplex(t=50) & 0.81±0.05 & 0.26±0.05 & \underline{78.9} & \textbf{70.8} \\
AnoDDPM~\cite{ho2020denoising} with Gaussian(t=250)  & \textbf{0.88±0.01} & \textbf{0.05±0.01} & 73.0 & {63.8}\\
AnoDDPM~\cite{ho2020denoising} with Gaussian(t=300) & 0.87±0.02 & \textbf{0.05±0.01}& {73.5} & 57.4 \\
\hline
AutoDDPM~\cite{bercea2023mask} with Gaussian(t=300)  & \textbf{0.88 $\pm$ 0.02} & \textbf{0.05 $\pm$ 0.02}& \textbf{79.8} & \underline{66.6} \\

\bottomrule
\end{tabular}
\label{tab:summary_classification}
\end{table}

\begin{figure}[!tb]
  \centering
    \subfloat[Healthy][Healthy example.]{\includegraphics[height=0.38\textheight]{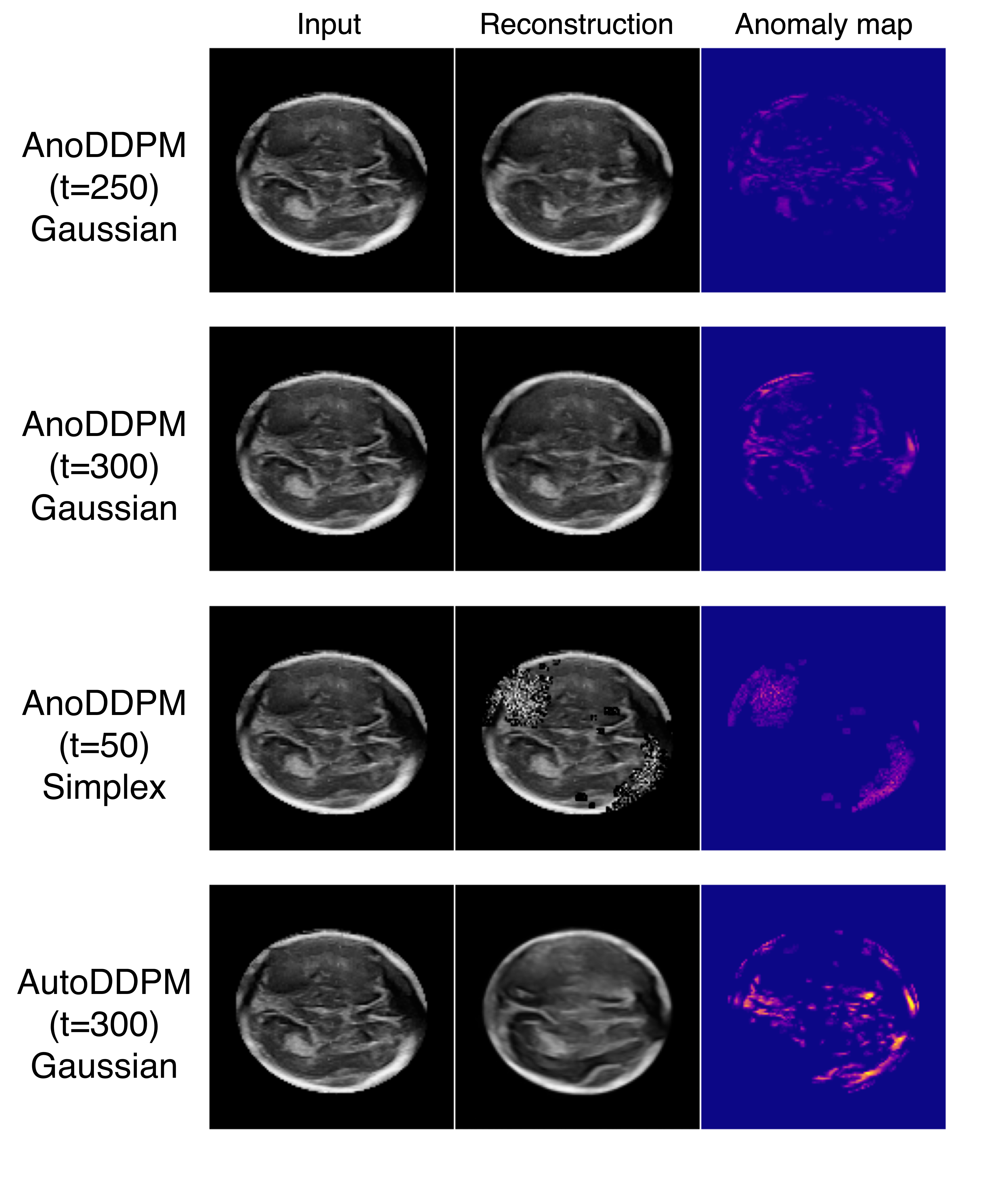}}\label{fig:best_healthy}
  \subfloat[Pathology][Pathological example.]{\includegraphics[height=0.38\textheight]{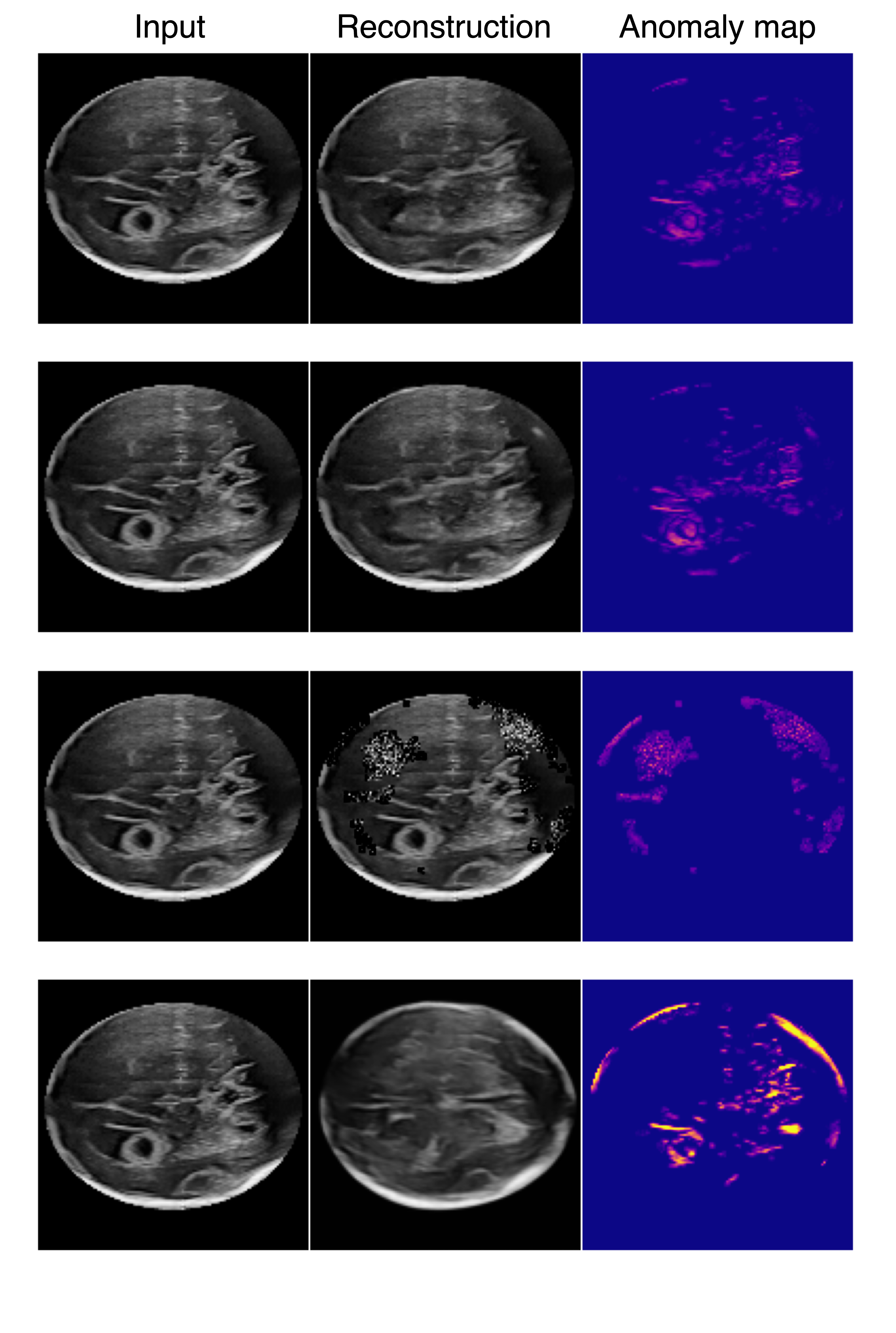}}\label{fig:best_pathology}
  \caption{Anomaly detection performance. From left to right we show the original image, the generated reconstruction, and the anomaly map.}
  \label{fig:best_results}
\end{figure}
We assessed the performance of AnoDDPM employing both Gaussian and Simplex Noise, alongside AutoDDPM, as detailed in~\cite{bercea2023mask}. The evaluation focused on various noise levels ($t={50, 100, 150, 200, 250, 300}$), analyzing the model's robustness under different conditions.

The classification outcomes are succinctly summarized in Table \ref{tab:classification}, with the most effective noise variations and types for both AnoDDPM and AutoDDPM highlighted. Corresponding visual results are depicted in Fig. \ref{fig:best_results}. For a thorough examination of all noise levels and types, including metrics on reconstruction errors and accuracy, readers are directed to the supplementary materials, which provide an extensive overview of all conducted experiments.

\begin{figure}[!tb]
    \centering
    \includegraphics[width=\linewidth]{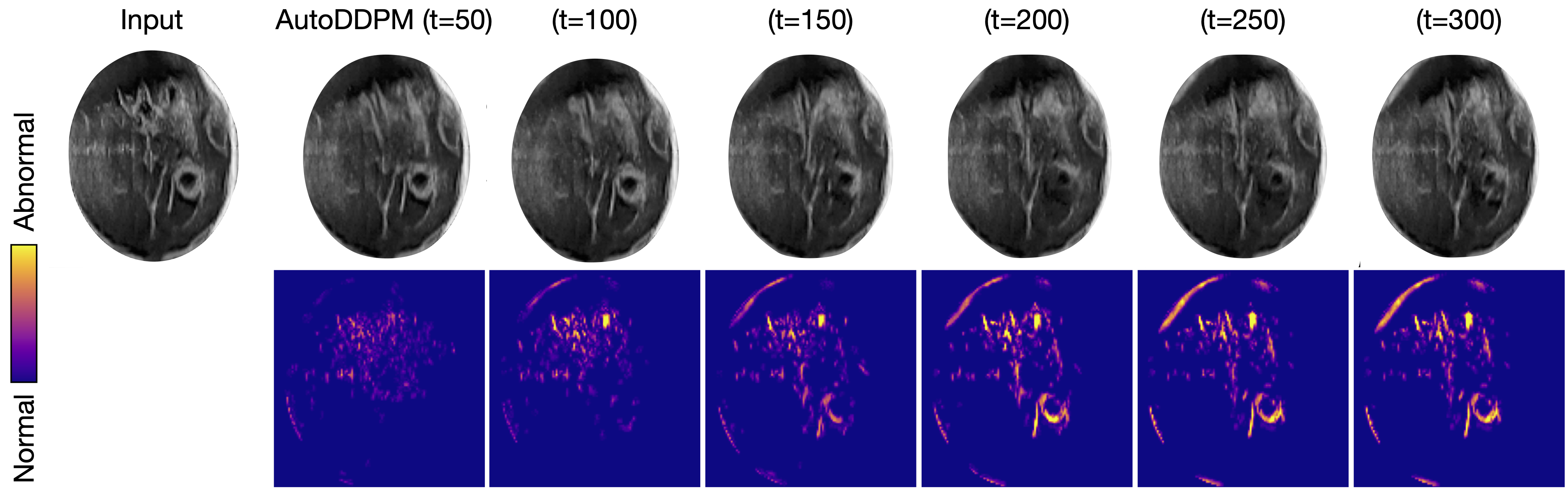}
    \caption{AutoDDPM~\cite{bercea2023mask} performance on pathology under multiple noise levels $t$.}
    \label{fig:auto_noise}
\end{figure}

\noindent\textbf{Influence of Noise Type and Levels.}
We evaluated AnoDDPM using Gaussian and Simplex noise at various noise levels, as summarized in Tables 2 and 3 in the supplementary materials. While Gaussian noise at a level of $t = 50$ yields the best performance in terms of reconstruction for healthy images (MAE of 0.013, SSIM of 0.898), it demonstrates limitations in anomaly detection with modest scores (67.5 AUPRC and 43.5 AUROC). This reflects a typical challenge in medical imaging: achieving minimal alteration in input images often compromises the effectiveness of anomaly detection. In contrast, higher noise levels ($t \geq 250$) show improved anomaly detection results, reaching up to 73 AUPRC and 63.8 AUROC, despite the compromise in image quality metrics. This observation aligns with the 'noise paradox' discussed in.\cite{bercea2023mask}, illustrating the complex trade-off between achieving optimal reconstruction of healthy images and effective anomaly detection in pathological cases. This paradox can also be observed in Fig.\ref{fig:auto_noise}, where higher noise levels enable the detection of the pathology but introduce false positive detections.

At a noise level of $t = 50$, AnoDDPM with Simplex noise achieves its most effective reconstruction capabilities for healthy images, recording MAE values of 0.034 and SSIM scores of 0.808. These significantly lag behind the results obtained with Gaussian noise, highlighting a substantial disparity in denoising effectiveness. As noise levels increase, both MAE and SSIM metrics for Simplex noise exhibit gradual deterioration, e.g., for $t=300$, the MAE increases to 0.045, while SSIM scores decrease to 0.741. In terms of anomaly detection capabilities, Simplex noise at $t = 50$ delivers the most impressive results, achieving an AUPRC of 78.9 and an AUROC of 70.8. These figures are the highest among all configurations and noise types tested, underscoring the superior ability of Simplex noise to detect anomalies under specific conditions, despite its lower reconstruction performance.

Figure \ref{fig:best_results} illustrates that Gaussian noise models achieve high-quality reconstructions and effectively highlight anomalies in fetal brain US images. In contrast, Simplex noise introduces noticeable noise artifacts on both healthy and pathological scans, though it remains effective in anomaly detection.\\

\noindent\textbf{Influence of Anomaly Maps.}
We analyzed the impact of different anomaly maps — MAE, LPIPS~\cite{bercea2023generalizing}, and their combination (MAE*LPIPS) — on anomaly detection performance. Detailed results are shown in Table 4 and Figures 1 and 2 in the supplementary materials. The combined anomaly maps yield the best performance overall, proving to be more precise in balancing the identification of pathological regions and reducing false positive detections.


%% file: sections/6_conclusion.tex
\section{Discussion}
Our study into the automatic anomaly detection pipeline for fetal brain US using DDPMs demonstrates effective pseudo-healthy synthesis and precise anomaly localization. Notably, newer diffusion models such as AutoDDPM~\cite{bercea2023mask} show significant promise in enhancing these capabilities. However, the performance of Simplex noise raises specific concerns; its interaction with the inherent noise characteristic of US images suggests it may not be ideally suited for these applications. This behavior necessitates further analysis to fully understand and mitigate adverse effects caused by this type of artificial noise.

Although the results are promising and the dataset was large enough for the models to be able to capture the underlying normative distribution, the statistical validation of these findings requires more extensive datasets. This project is part of an ongoing initiative within the clinic, with continuous efforts to collect more US images to enrich our dataset. Additionally, assessing the clinical utility of this automated screening for pathologies is vital. Evaluating how well our pipeline supports operators and sonographers in detecting and diagnosing conditions will help determine the practical benefits of integrating this technology into everyday clinical practice.

Future research directions will focus on evaluating our pipeline on US video data to simulate real clinical environments more accurately. This includes capturing the dynamic aspects of fetal movements and variations in US probe positioning, which can significantly impact image quality and diagnostic accuracy. Moreover, expanding the diversity of the dataset with images from different US machines and settings will help improve the generalizability and robustness of the proposed methods.

\section{Conclusion}
In this work, we introduced a novel framework to enable fetal brain anomaly detection on US scans. We designed the framework to automatically reduce the noise and randomness inherent in the US images, allowing models to learn the normative distribution effectively. For the first time, we applied diffusion-based models to automatically processed fetal brain US scans and evaluated their performance under different noise types and levels. Our results indicate that diffusion models, particularly AutoDDPM, hold significant potential for improving the accuracy and reliability of fetal brain anomaly detection in clinical settings.

%% file: sections/99_apendix.tex
\section{Appendix}
\begin{table}[ht]
\centering
\begin{tabular}{@{}l|l|l|l|l@{}}
\hline
& \multicolumn{2}{c}{Rec. error - MAE} & \multicolumn{2}{c}{Rec. accuracy - SSIM} \\
\hline
& Pathology & Healthy $\downarrow$ & Pathology & Healthy $\uparrow$ \\
\hline
Gaussian(t=50)  & 0.013 $\pm$ 0.002 & \textbf{0.013 $\pm$ 0.001} & 0.889 $\pm$ 0.019 & \textbf{0.898 $\pm$ 0.007} \\
Gaussian(t=100) & 0.014 $\pm$ 0.002 & 0.014 $\pm$ 0.001 & 0.882 $\pm$ 0.018 & 0.895 $\pm$ 0.011 \\
Gaussian(t=150) & 0.015 $\pm$ 0.002 & 0.015 $\pm$ 0.001 & 0.875 $\pm$ 0.014 & 0.887 $\pm$ 0.010 \\
Gaussian(t=200) & 0.016 $\pm$ 0.002 & 0.016 $\pm$ 0.001 & 0.869 $\pm$ 0.016 & 0.879 $\pm$ 0.010 \\
Gaussian(t=250) & 0.017 $\pm$ 0.002 & 0.016 $\pm$ 0.001 & 0.863 $\pm$ 0.014 & 0.876 $\pm$ 0.011 \\
Gaussian(t=300) & 0.018 $\pm$ 0.001 & 0.018 $\pm$ 0.001 & 0.856 $\pm$ 0.013 & 0.867 $\pm$ 0.017 \\
\hline
Simplex(t=50)  & 0.043 $\pm$ 0.006 & \textbf{0.034 $\pm$ 0.006} & 0.743 $\pm$ 0.040 & \textbf{0.808 $\pm$ 0.047} \\
Simplex(t=100) & 0.047 $\pm$ 0.005 & 0.046 $\pm$ 0.006 & 0.734 $\pm$ 0.026 & 0.749 $\pm$ 0.020 \\
Simplex(t=150) & 0.050 $\pm$ 0.009 & 0.047 $\pm$ 0.004 & 0.718 $\pm$ 0.038 & 0.741 $\pm$ 0.020 \\
Simplex(t=200) & 0.051 $\pm$ 0.008 & 0.048 $\pm$ 0.007 & 0.715 $\pm$ 0.038 & 0.731 $\pm$ 0.036 \\
Simplex(t=250) & 0.049 $\pm$ 0.007 & 0.046 $\pm$ 0.005 & 0.721 $\pm$ 0.037 & 0.741 $\pm$ 0.027 \\
Simplex(t=300) & 0.048 $\pm$ 0.006 & 0.045 $\pm$ 0.006 & 0.728 $\pm$ 0.032 & 0.741 $\pm$ 0.026 \\
\hline
AutoDDPM(t=50)  & 0.017 $\pm$ 0.003 & 
0.017 $\pm$ 0.002 & 0.864 $\pm$ 0.016 & 0.883 $\pm$ 0.015 \\
AutoDDPM(t=100) & 0.015 $\pm$ 0.003 & \textbf{0.014 $\pm$ 0.002} & 0.882 $\pm$ 0.016 & \textbf{0.894 $\pm$ 0.013} \\
AutoDDPM(t=150) & 0.016 $\pm$ 0.003 & 0.015 $\pm$ 0.002 & 0.873 $\pm$ 0.016 & 0.888 $\pm$ 0.012 \\
AutoDDPM(t=200) & 0.016 $\pm$ 0.003 & 0.016 $\pm$ 0.002 & 0.869 $\pm$ 0.017 & 0.887 $\pm$ 0.011 \\
AutoDDPM(t=250) & 0.017 $\pm$ 0.003 & 0.017 $\pm$ 0.002 & 0.864 $\pm$ 0.016 & 0.883 $\pm$ 0.015 \\
AutoDDPM(t=300) & 0.018 $\pm$ 0.003 & 0.018 $\pm$ 0.002 & 0.862 $\pm$ 0.015 & 0.878 $\pm$ 0.020 \\
\hline
\end{tabular}
\caption{Reconstruction errors using MAE and reconstruction accuracies using SSIM for AnoDDPM with Gaussian or Simplex noise types, and AutoDDPM with Gaussian noise type at various noise levels.}
\label{tab:reconstruction_error_ssim}
\end{table}

\begin{table}[ht]
\centering
\begin{tabular}{@{}l|l|ll@{}}
\hline
& \multicolumn{2}{c}{Rec. accuracy - LPIPS} \\
\hline
& Pathology & Healthy $\downarrow$ \\
\hline
Gaussian(t=50) & 0.042 $\pm$ 0.012 & \textbf{0.037 $\pm$ 0.007} \\
Gaussian(t=100) & 0.045 $\pm$ 0.010 & 0.039 $\pm$ 0.006 \\
Gaussian(t=150) & 0.050 $\pm$ 0.011 & 0.044 $\pm$ 0.007 \\
Gaussian(t=200) & 0.055 $\pm$ 0.016 & 0.047 $\pm$ 0.011 \\
Gaussian(t=250) & 0.058 $\pm$ 0.015 & 0.048 $\pm$ 0.011 \\
Gaussian(t=300) & 0.064 $\pm$ 0.017 & 0.053 $\pm$ 0.010 \\
\hline
Simplex(t=50)  & 0.347 $\pm$ 0.054 & \textbf{0.263 $\pm$ 0.048} \\
Simplex(t=100) & 0.379 $\pm$ 0.050 & 0.353 $\pm$ 0.036 \\
Simplex(t=150) & 0.400 $\pm$ 0.057 & 0.388 $\pm$ 0.057\\
Simplex(t=200)& 0.400 $\pm$ 0.042 & 0.387 $\pm$ 0.046 \\
Simplex(t=250) & 0.373 $\pm$ 0.035 & 0.371 $\pm$ 0.041 \\
Simplex(t=300) & 0.375 $\pm$ 0.033 & 0.365 $\pm$ 0.047 \\
\hline
AutoDDPM(t=50)  & 0.062 $\pm$ 0.017 & 0.044 $\pm$ 0.012 \\
AutoDDPM(t=100) & 0.046 $\pm$ 0.011 & \textbf{0.039 $\pm$ 0.007} \\
AutoDDPM(t=150) & 0.053 $\pm$ 0.015 & 0.044 $\pm$ 0.010\\
AutoDDPM(t=200)& 0.058 $\pm$ 0.016 & 0.046 $\pm$ 0.010 \\
AutoDDPM(t=250) & 0.061 $\pm$ 0.017 & 0.044 $\pm$ 0.012 \\
AutoDDPM(t=300) & 0.066 $\pm$ 0.015 & 0.047 $\pm$ 0.016 \\
\hline
\end{tabular}
\caption{Reconstruction accuracy using LPIPS for AnoDDPM with Gaussian or Simplex noise types, and AutoDDPM with Gaussian noise type at various noise levels.}
\label{tab:reconstruction2}

\end{table}
\begin{table}[ht]
\centering
\begin{tabular}{l|c|c}
\hline
& \multicolumn{2}{c}{MAE*LPIPS}\\
\hline
 & AURPC $\uparrow$ & AUROC $\uparrow$ \\
\hline
Gaussian(t=50) & 67.5 & 43.5 \\
Gaussian(t=100) & 62.5 & 48.6 \\
Gaussian(t=150) & 67.9 & 51.8 \\
Gaussian(t=200) & 66.9 & 51.8 \\
Gaussian(t=250) & 73.0 & \textbf{63.8} \\
Gaussian(t=300) & \textbf{73.5} & 57.4 \\
\hline
Simplex(t=50) & \textbf{78.9} & \textbf{70.8} \\
Simplex(t=100) & 69.1 & 50.0 \\
Simplex(t=150) & 59.0 & 33.3 \\
Simplex(t=200) & 65.2 & 46.2 \\
Simplex(t=250) & 57.6 & 45.8 \\
Simplex(t=300) & 65.9 & 49.0 \\
\hline
AutoDDPM(t=50) & 71.1 & 54.6 \\
AutoDDPM(t=100) & 73.7 & 63.4 \\
AutoDDPM(t=150) & 73.1 & 62.5\\
AutoDDPM(t=200) & 77.6 & 65.7 \\
AutoDDPM(t=250) & 69.5 & 55.5 \\
AutoDDPM(t=300) & \textbf{79.8} & \textbf{66.6} \\
\hline
\end{tabular}
\caption{Alternative classification measures using MAE*LPIPS anomaly maps: AUPRC and AUROC for unsupervised anomaly detection utilizing AnoDDPM with Gaussian or Simplex noise types, and AutoDDPM with Gaussian noise type at various noise levels.}
\label{tab:classification}
\end{table}

\begin{figure}[!tbp]
  \centering
  \subfloat[Pathology][Pathological example.]{\includegraphics[height=0.5\textheight]{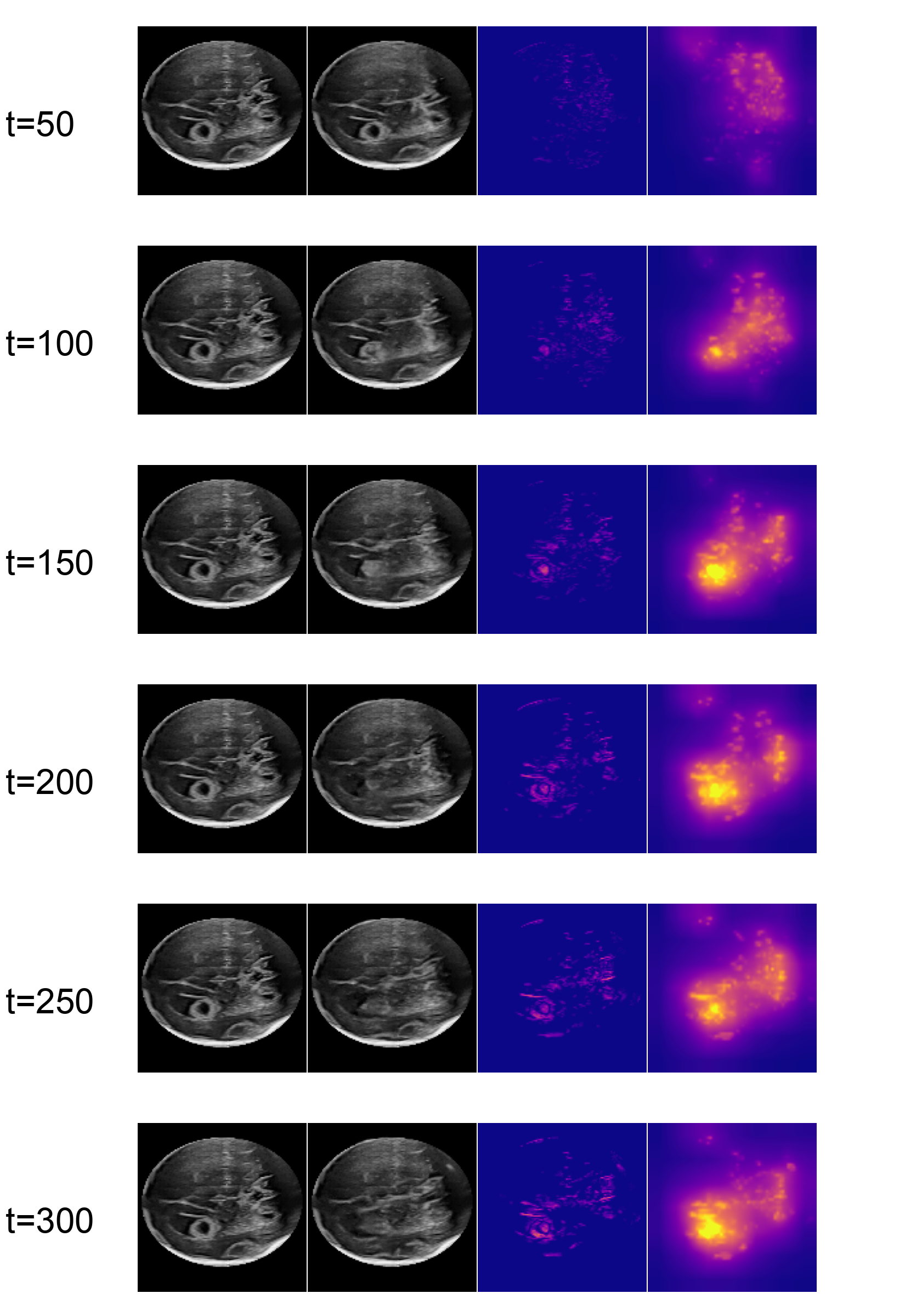}\label{fig:gaussian1}}
  \subfloat[Healthy][Healthy example.]{\includegraphics[height=0.5\textheight]{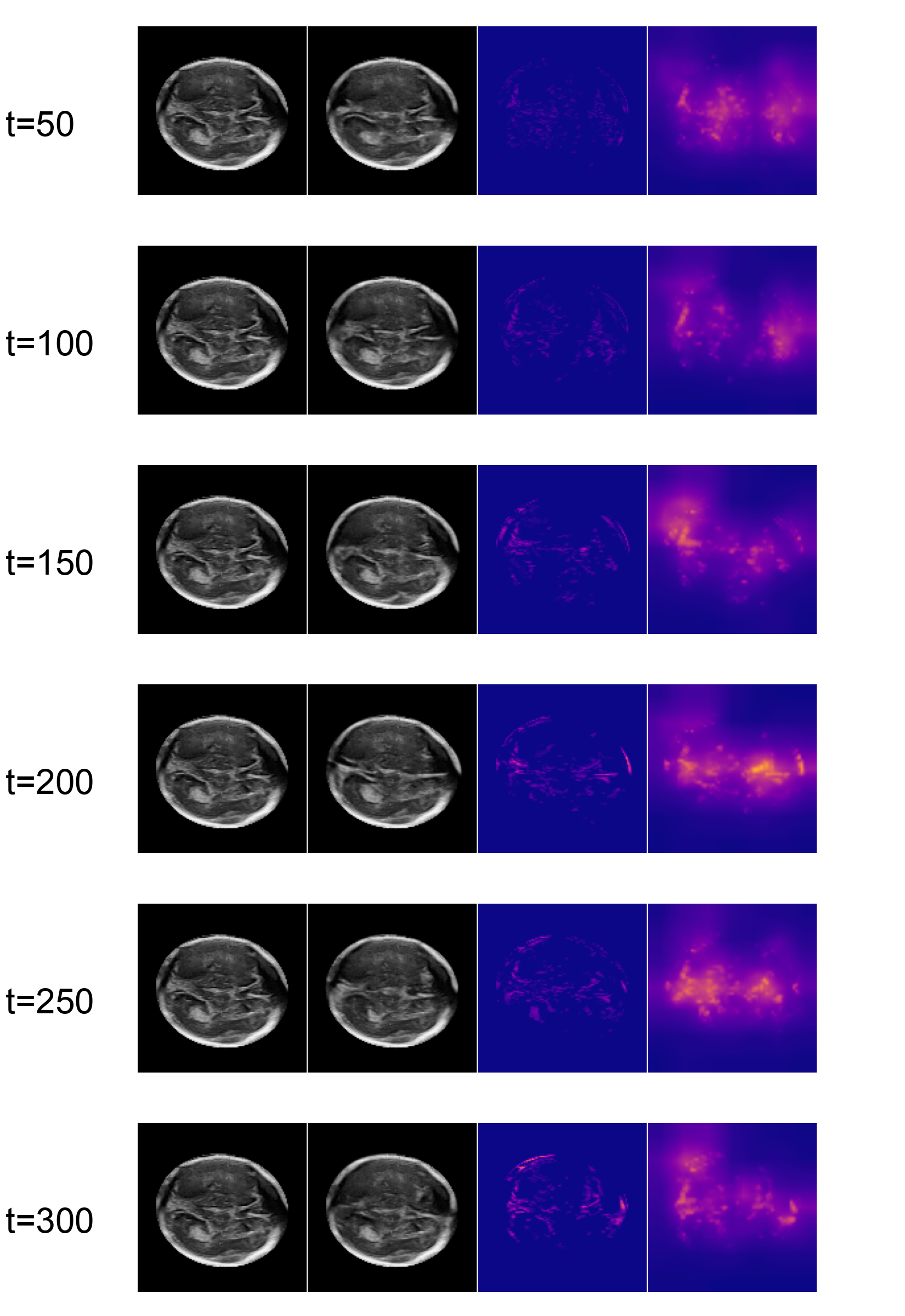}\label{fig:gaussian2}}
  \caption{Visualization of reconstructions and anomaly maps using Gaussian noise for different noise levels. Each image contains 4 images in the corresponding order: the original image, the generated reconstruction, the anomaly map using MAE*LPIPS loss, the anomaly map using LPIPS loss.}
  \label{fig:gaussian}
\end{figure}

\begin{figure}[!tbp]
  \centering
  \subfloat[Pathology][Pathological example.]{\includegraphics[height=0.5\textheight]{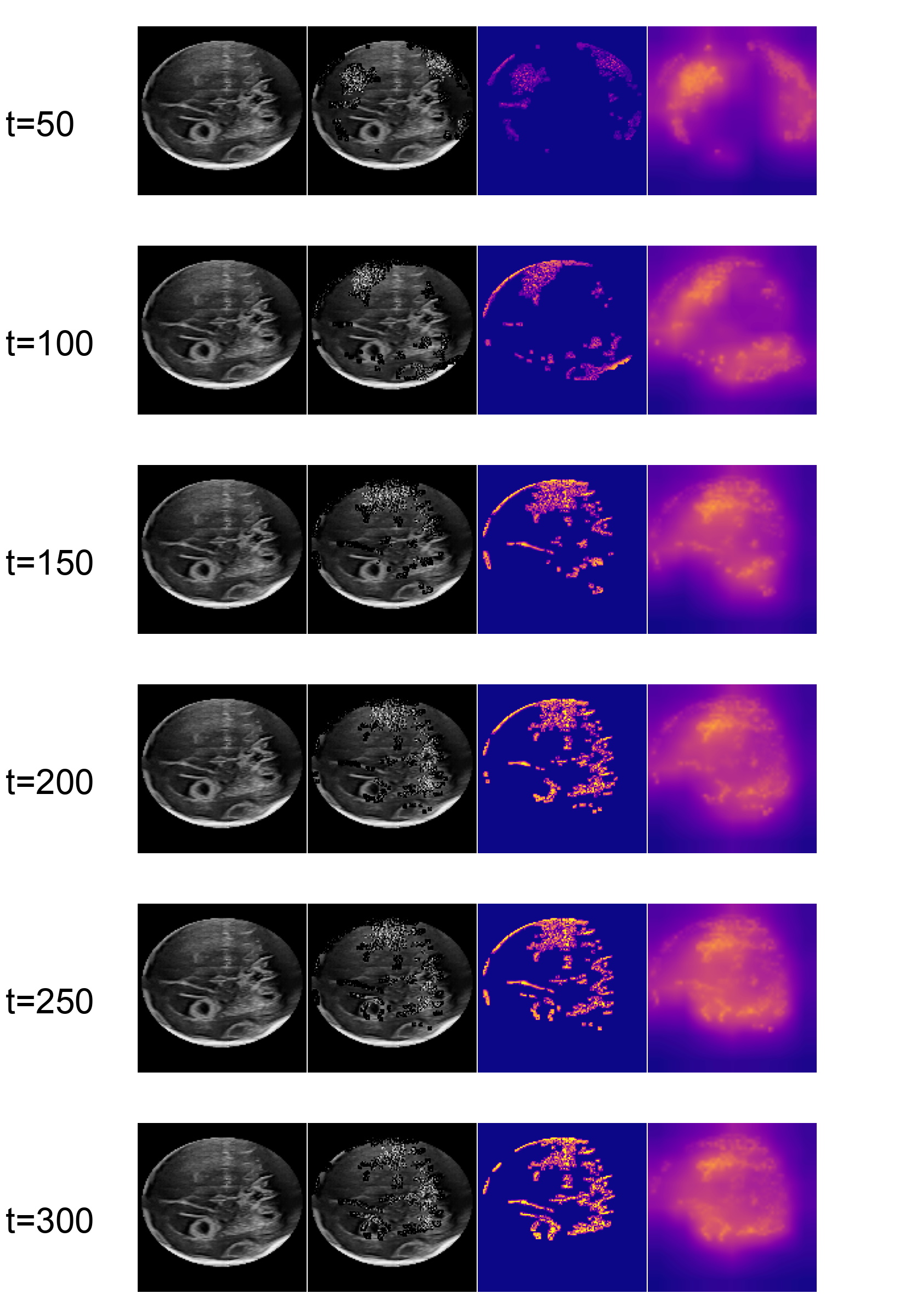}\label{fig:simplex1}}
  \subfloat[Healthy][Healthy example.]{\includegraphics[height=0.5\textheight]{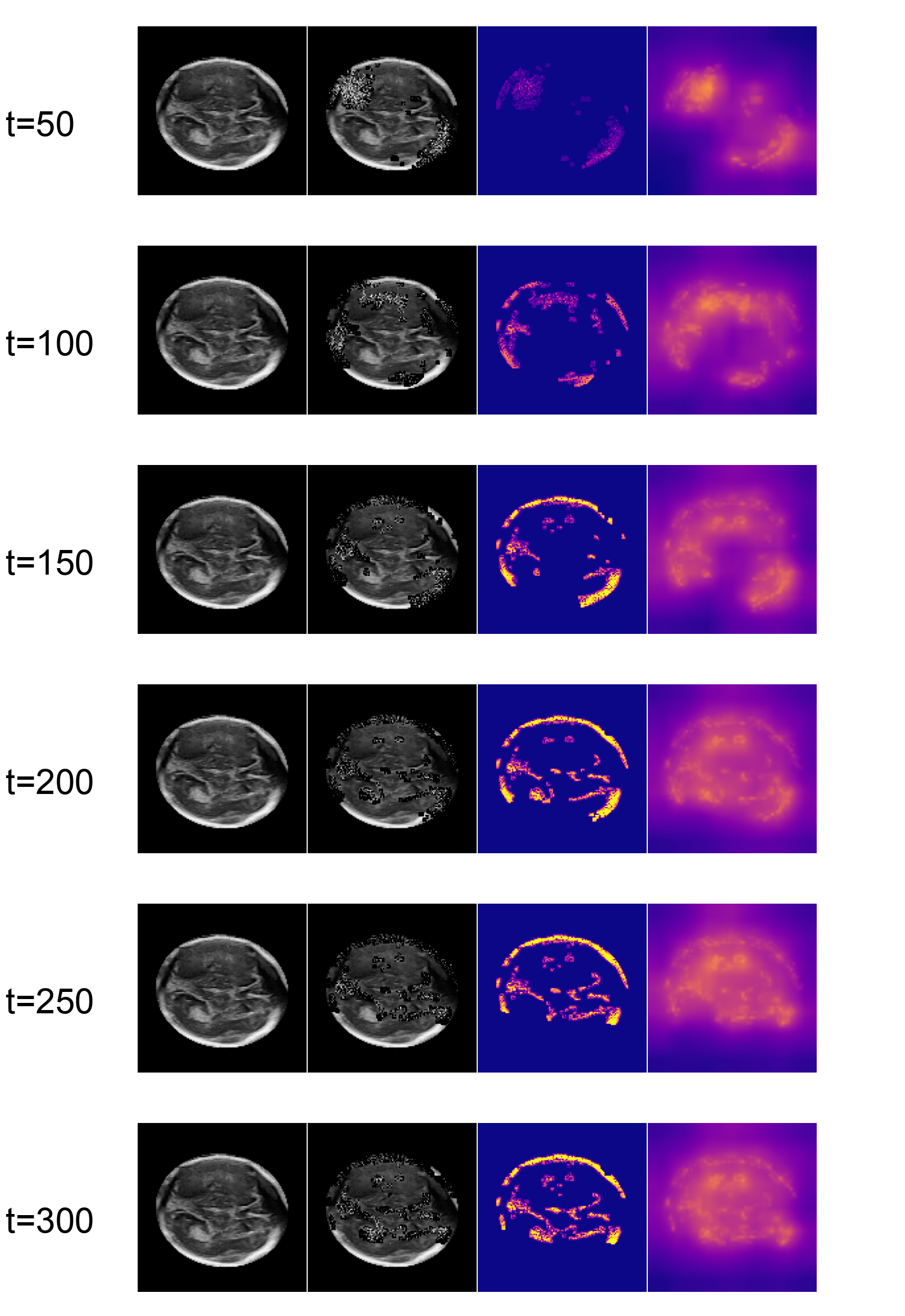}\label{fig:simplex2}}
  \caption{Visualization of reconstructions and anomaly maps using Simplex noise for different noise levels. Each image contains 4 images in the corresponding order: the original image, the generated reconstruction, the anomaly map using MAE*LPIPS loss, the anomaly map using LPIPS loss.}
  \label{fig:simplex}
\end{figure}

\begin{figure}[!tbp]
  \centering
  \subfloat[Pathology][Pathological example.]{\includegraphics[height=0.5\textheight]{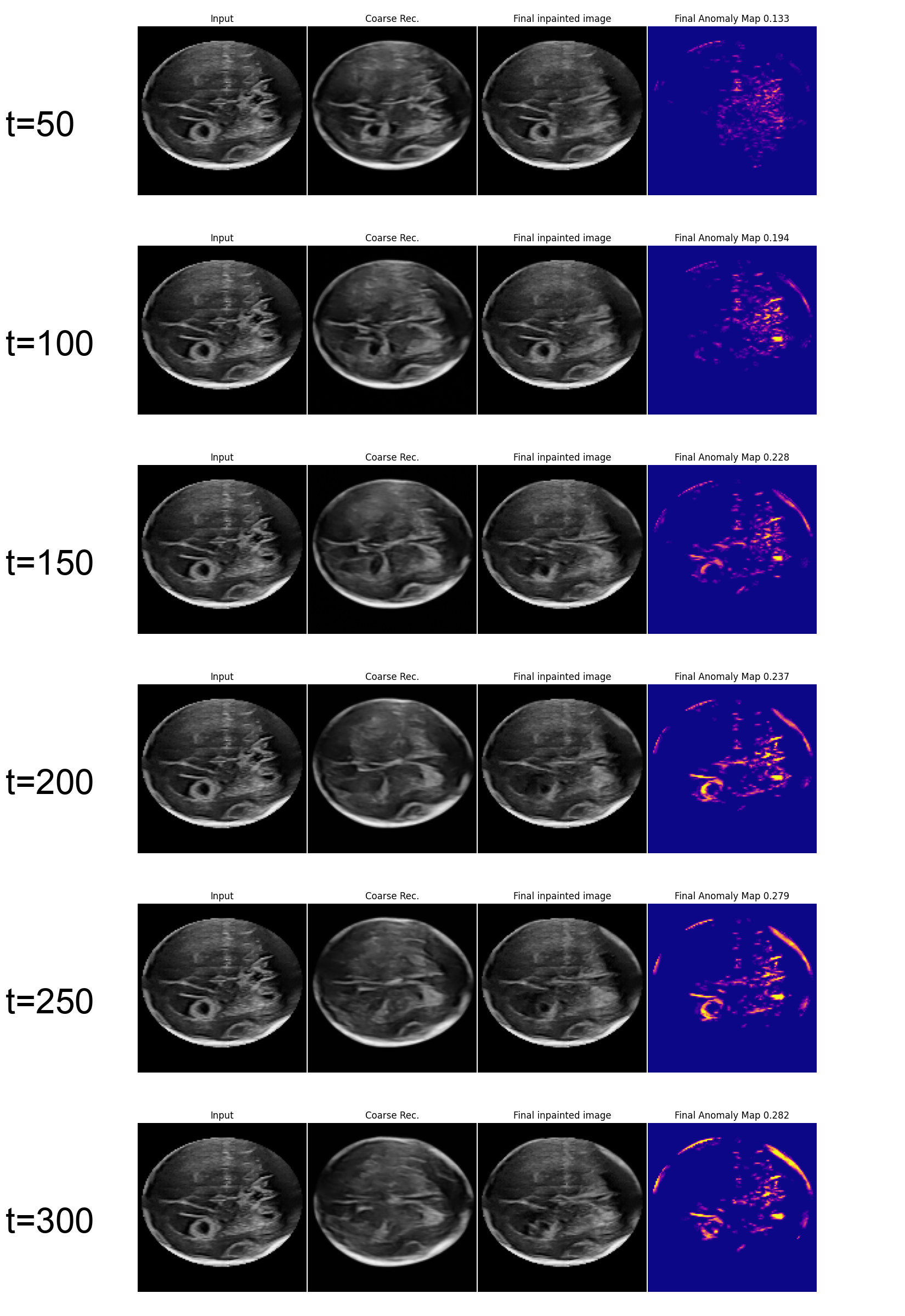}\label{fig:simplex1}}
  \subfloat[Healthy][Healthy example.]{\includegraphics[height=0.5\textheight]{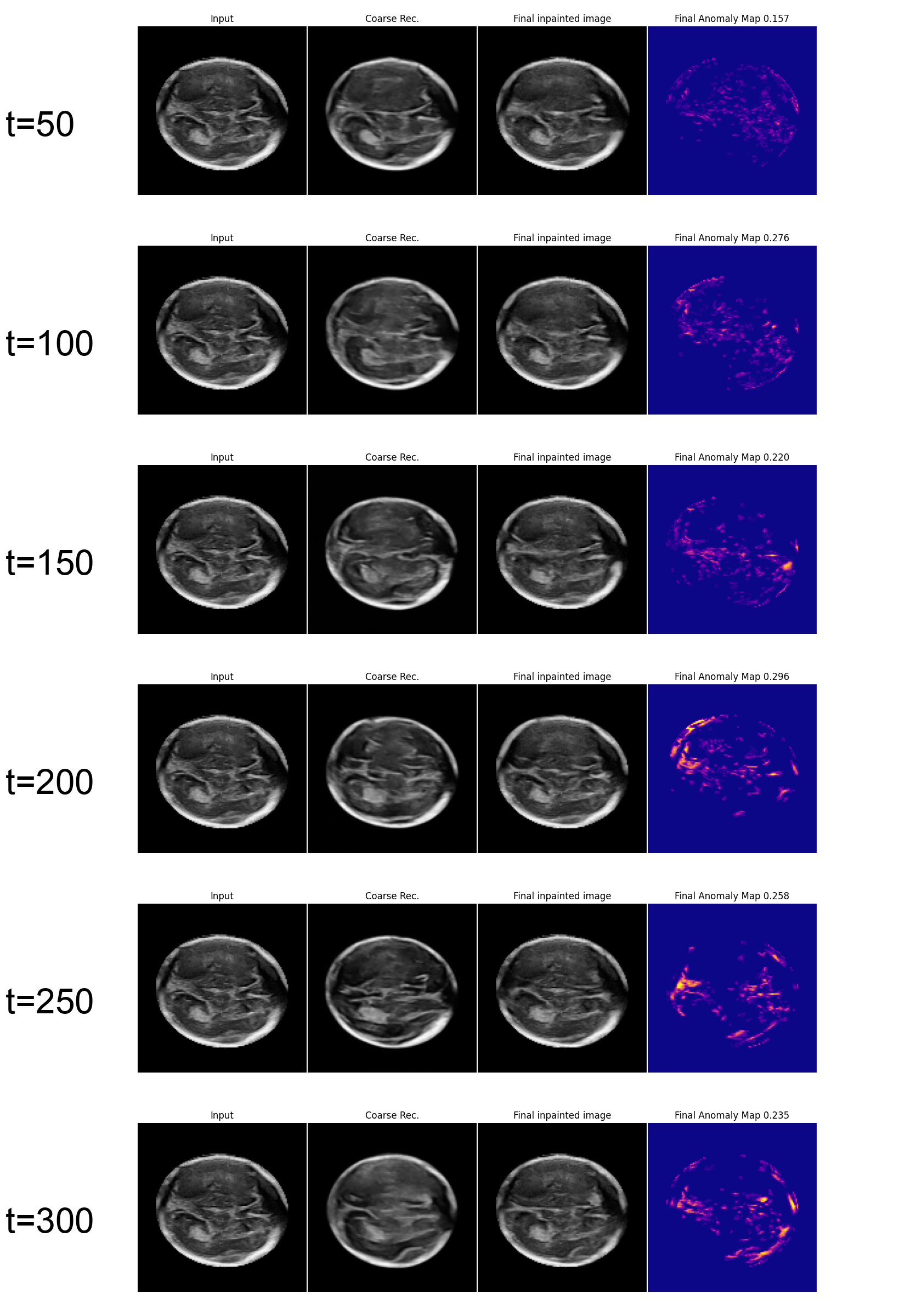}\label{fig:simplex2}}
  \caption{Visualization of reconstructions and anomaly maps using AutoDDPM with Gaussian noise for different noise levels. Each image contains 4 images in the corresponding order: the original image, the generated coarse reconstruction, the final reconstruction, and the anomaly map using MAE*LPIPS loss.}
  \label{fig:autoddpm}
\end{figure}

%% file: 0_main.bbl
\begin{thebibliography}{10}
\providecommand{\url}[1]{\texttt{#1}}
\providecommand{\urlprefix}{URL }
\providecommand{\doi}[1]{https://doi.org/#1}

\bibitem{akcay2019ganomaly}
Akcay, S., Atapour-Abarghouei, A., Breckon, T.P.: Ganomaly: Semi-supervised
  anomaly detection via adversarial training. In: Computer Vision--ACCV 2018:
  14th Asian Conference on Computer Vision, Perth, Australia, December 2--6,
  2018, Revised Selected Papers, Part III 14. pp. 622--637. Springer (2019)

\bibitem{baumgartner2017sononet}
Baumgartner, C.F., Kamnitsas, K., Matthew, J., Fletcher, T.P., Smith, S., Koch,
  L.M., Kainz, B., Rueckert, D.: Sononet: Real-time detection and localisation
  of fetal standard scan planes in freehand ultrasound (2017)

\bibitem{behrendt2023patched}
Behrendt, F., Bhattacharya, D., Kr{\"u}ger, J., Opfer, R., Schlaefer, A.:
  Patched diffusion models for unsupervised anomaly detection in brain mri.
  International Conference on Medical Imaging with Deep Learning  (2023)

\bibitem{bercea2023mask}
Bercea, C.I., Neumayr, M., Rueckert, D., Schnabel, J.A.: Mask, stitch, and
  re-sample: Enhancing robustness and generalizability in anomaly detection
  through automatic diffusion models (2023)

\bibitem{bercea2023aes}
Bercea, C.I., Rueckert, D., Schnabel, J.A.: What do aes learn? challenging
  common assumptions in unsupervised anomaly detection. In: International
  Conference on Medical Image Computing and Computer-Assisted Intervention. pp.
  304--314. Springer (2023)

\bibitem{bercea2023generalizing}
Bercea, C.I., Wiestler, B., Rueckert, D., Schnabel, J.A.: Generalizing
  unsupervised anomaly detection: towards unbiased pathology screening. In:
  Medical Imaging with Deep Learning (2023)

\bibitem{biometrics_2}
Budd, S., Sinclair, M., Khanal, B., Matthew, J., Lloyd, D., Gomez, A.,
  Toussaint, N., Robinson, E.C., Kainz, B.: Confident head circumference
  measurement from ultrasound with real-time feedback for sonographers. In:
  Shen, D., Liu, T., Peters, T.M., Staib, L.H., Essert, C., Zhou, S., Yap,
  P.T., Khan, A. (eds.) Medical Image Computing and Computer Assisted
  Intervention -- MICCAI 2019. pp. 683--691. Springer International Publishing,
  Cham (2019)

\bibitem{burgos_artizzu_2020_3904280}
Burgos-Artizzu, X.P., Coronado-Gutierrez, D., Valenzuela-Alcaraz, B.,
  Bonet-Carne, E., Eixarch, E., Crispi, F., Gratacós, E.: {FETAL\_PLANES\_DB:
  Common maternal-fetal ultrasound images} (Jun 2020).
  \doi{10.5281/zenodo.3904280}, \url{https://doi.org/10.5281/zenodo.3904280}

\bibitem{chen2020unsupervised}
Chen, X., You, S., Tezcan, K.C., Konukoglu, E.: Unsupervised lesion detection
  via image restoration with a normative prior. Medical Image Analysis
  \textbf{64} (2020)

\bibitem{kainz_fetal_heart}
Chotzoglou, E., Day, T., Tan, J., Matthew, J., Lloyd, D., Razavi, R., Simpson,
  J., Kainz, B., et~al.: Learning normal appearance for fetal anomaly
  screening: Application to the unsupervised detection of hypoplastic left
  heart syndrome. Machine Learning for Biomedical Imaging  \textbf{1}(September
  2021 issue),  1--25 (2021)

\bibitem{Fiorentino_2023}
Fiorentino, M.C., Villani, F.P., Di~Cosmo, M., Frontoni, E., Moccia, S.: A
  review on deep-learning algorithms for fetal ultrasound-image analysis.
  Medical Image Analysis  \textbf{83},  102629 (Jan 2023).
  \doi{10.1016/j.media.2022.102629},
  \url{http://dx.doi.org/10.1016/j.media.2022.102629}

\bibitem{DBLP:journals/corr/abs-2006-11239}
Ho, J., Jain, A., Abbeel, P.: Denoising diffusion probabilistic models. CoRR
  \textbf{abs/2006.11239} (2020), \url{https://arxiv.org/abs/2006.11239}

\bibitem{ho2020denoising}
Ho, J., Jain, A., Abbeel, P.: Denoising diffusion probabilistic models (2020)

\bibitem{10.3389/frai.2022.832485}
Lo, J., Lim, A., Wagner, M.W., Ertl-Wagner, B., Sussman, D.: Fetal organ
  anomaly classification network for identifying organ anomalies in fetal mri.
  Frontiers in Artificial Intelligence  \textbf{5} (2022).
  \doi{10.3389/frai.2022.832485},
  \url{https://www.frontiersin.org/articles/10.3389/frai.2022.832485}

\bibitem{Luo_2023}
Luo, G., Xie, W., Gao, R., Zheng, T., Chen, L., Sun, H.: Unsupervised anomaly
  detection in brain mri: Learning abstract distribution from massive healthy
  brains. Computers in Biology and Medicine  \textbf{154},  106610 (2023).
  \doi{https://doi.org/10.1016/j.compbiomed.2023.106610},
  \url{https://www.sciencedirect.com/science/article/pii/S0010482523000756}

\bibitem{Meng_2019}
Meng, L., Zhao, D., Yang, Z., Wang, B.: Automatic display of fetal brain planes
  and automatic measurements of fetal brain parameters by transabdominal
  three‐dimensional ultrasound. Journal of Clinical Ultrasound  \textbf{48}
  (07 2019). \doi{10.1002/jcu.22762}

\bibitem{Pang_2021}
Pang, G., Shen, C., Cao, L., Hengel, A.V.D.: Deep learning for anomaly
  detection: A review. ACM Computing Surveys  \textbf{54}(2),  1–38 (Mar
  2021). \doi{10.1145/3439950}, \url{http://dx.doi.org/10.1145/3439950}

\bibitem{PrenatalImaging}
Reddy, U., Filly, R., Copel, J.: Prenatal imaging: Ultrasonography and magnetic
  resonance imaging. Obstetrics and gynecology  \textbf{112},  145--57 (08
  2008). \doi{10.1097/01.AOG.0000318871.95090.d9}

\bibitem{ronneberger2015unet}
Ronneberger, O., Fischer, P., Brox, T.: U-net: Convolutional networks for
  biomedical image segmentation (2015)

\bibitem{schlegl2019fanogan}
Schlegl, T., Seeböck, P., Waldstein, S.M., Langs, G., Schmidt-Erfurth, U.:
  f-{A}no{GAN}: Fast unsupervised anomaly detection with generative adversarial
  networks. Medical Image Analysis  \textbf{54},  30--44 (2019)

\bibitem{biometrics}
Sinclair, M., Baumgartner, C.F., Matthew, J., Bai, W., Martinez, J.C., Li, Y.,
  Smith, S., Knight, C.L., Kainz, B., Hajnal, J., King, A.P., Rueckert, D.:
  Human-level performance on automatic head biometrics in fetal ultrasound
  using fully convolutional neural networks. In: 2018 40th Annual International
  Conference of the IEEE Engineering in Medicine and Biology Society (EMBC).
  pp. 714--717 (2018). \doi{10.1109/EMBC.2018.8512278}

\bibitem{Tschuchnig_2022}
Tschuchnig, M.E., Gadermayr, M.: Anomaly Detection in Medical Imaging - A Mini
  Review, p. 33–38. Springer Fachmedien Wiesbaden (2022).
  \doi{10.1007/978-3-658-36295-9_5},
  \url{http://dx.doi.org/10.1007/978-3-658-36295-9_5}

\bibitem{wolleb2022diffusion}
Wolleb, J., Bieder, F., Sandk{\"u}hler, R., Cattin, P.C.: Diffusion models for
  medical anomaly detection. Medical Image Computing and Computer Assisted
  Intervention pp. 35--45 (2022)

\bibitem{Wyatt_2022_CVPR}
Wyatt, J., Leach, A., Schmon, S.M., Willcocks, C.G.: Anoddpm: Anomaly detection
  with denoising diffusion probabilistic models using simplex noise. In:
  Proceedings of the IEEE/CVF Conference on Computer Vision and Pattern
  Recognition (CVPR) Workshops. pp. 650--656 (06 2022)

\bibitem{xie_2020}
Xie, H.N., Wang, N., He, M., Zhang, L.H., Cai, H.M., Xian, J.B., Lin, M.F.,
  Zheng, J., Yang, Y.Z.: Using deep-learning algorithms to classify fetal brain
  ultrasound images as normal or abnormal. Ultrasound in Obstetrics \&
  Gynecology  \textbf{56}(4),  579--587 (2020).
  \doi{https://doi.org/10.1002/uog.21967},
  \url{https://obgyn.onlinelibrary.wiley.com/doi/abs/10.1002/uog.21967}

\bibitem{Zhang_2022}
Zhang, H., Guo, W., Zhang, S., Lu, H., Zhao, X.: Unsupervised deep anomaly
  detection for medical images using an improved adversarial autoencoder.
  Journal of Digital Imaging  \textbf{35} (01 2022).
  \doi{10.1007/s10278-021-00558-8}

\bibitem{zhang2018unreasonable}
Zhang, R., Isola, P., Efros, A.A., Shechtman, E., Wang, O.: The unreasonable
  effectiveness of deep features as a perceptual metric (2018)

\bibitem{zimmerer2019unsupervised}
Zimmerer, D., Isensee, F., Petersen, J., Kohl, S., Maier-Hein, K.: Unsupervised
  anomaly localization using variational auto-encoders (2019)

\end{thebibliography}
